\documentclass[12pt]{article}
\usepackage[font=small,labelfont=small]{caption}
\usepackage[section] {placeins}

\usepackage[papersize={8.5in,11in},hdivide={1in,*,1in},vdivide={1in,*,1in},dvips,mag=1000,nofoot]{geometry}
\usepackage{graphicx}
\usepackage{amsmath,amsfonts,amssymb}
\usepackage{natbib}
\bibpunct{(}{)}{;}{a}{,}{,}

\numberwithin{equation}{section}

\long\def\symbolfootnote[#1]#2{\begingroup
\def\thefootnote{\fnsymbol{footnote}}\footnote[#1]{#2}\endgroup}

\newcommand\solidrule[1][1cm]{\rule[0.5ex]{#1}{.4pt}}

\newcommand\dashedrule{\mbox{%
\solidrule[1mm]\hspace{1mm}\solidrule[1mm]\hspace{1mm}\solidrule[1mm]\hspace{1mm}\solidrule[1mm]\hspace{1mm}\solidrule[1mm]\hspace{1mm}}}

\begin{document}

\noindent \textbf{\large{Bayesian Benchmark Dose Analysis}}

\vspace{4 pt}
\renewcommand\thefootnote{\fnsymbol{footnote}}
\noindent \textbf{Qijun Fang}$^{\textrm{a}}$\symbolfootnote[2]{Email address: qijunf@email.arizona.edu; Corresponding author}, \textbf{Walter W.~Piegorsch}$^{\textrm{a,b}}$, and \textbf{Katherine Y.~Barnes}$^{\textrm{a,c}}$
\\
\noindent $^{\textrm{a}}$\textit{Program in Statistics}, $^{\textrm{b}}$\textit{BIO5 Institute}, and $^{\textrm{c}}$\textit{James E.~Rogers College of Law}\\
\textit{University of Arizona, Tucson, AZ, USA}\\


\noindent \textbf{Abstract}\\
\small
\noindent An important objective in environmental risk assessment is estimation of minimum exposure levels, called Benchmark Doses (BMDs) that induce a pre-specified Benchmark Response (BMR) in a target population. Established inferential approaches for BMD analysis typically involve one-sided, frequentist confidence limits, leading in practice to what are called Benchmark Dose Lower Limits (BMDLs). Appeal to Bayesian modeling and credible limits for building BMDLs is far less developed, however. Indeed, for the few existing forms of Bayesian BMDs, informative prior information is seldom incorporated. We develop reparameterized quantal-response models that explicitly describe the BMD as a target parameter.
Our goal is to obtain an improved estimation and calculation archetype for the BMD and for the BMDL, by employing quantifiable prior belief to represent parameter uncertainty in the statistical model.
Implementation is facilitated via a Monte Carlo-based adaptive Metropolis (AM) algorithm to approximate the posterior distribution. An example from environmental carcinogenicity testing illustrates the calculations.

\vspace{10 pt}


\section{Introduction}\label{sec:intro}
\subsection{\normalsize Benchmark Risk Analysis}\label{sec:BMD RA}
\indent \indent A primary objective in environmental risk assessment is characterization of the severity and likelihood of adverse outcomes caused by a hazardous agent \citep{ster08}.  The outcome could be death, cancer, mutation, damage caused by environmental or ecological hazards, etc.  In this context, the
\emph{risk function}, $R(d)$, is defined as the probability of exhibiting the adverse effect in a subject, object, or system exposed to a particular dose level, $d$, of the hazardous agent.
To illustrate, consider the following example.

\noindent \textit{Example 1. Benchmarking mammalian carcinogenicity of cumene.}\\
Cumene, the colloquial name for isopropylbenzene (C$_9$H$_{12}$), is a hydrocarbon solvent employed in the production of industrial compounds such as phenol and acetone.  Occupational and industrial exposures to cumene are common, so the U.S. National Toxicology Program (NTP) explored various forms of mammalian toxicity to the chemical \citep{tr542}. For example, Table \ref{tabl:data} displays quantal-response data on induction of lung tumors (alveolar/bronchiolar adenomas and carcinomas) by cumene in laboratory mice after chronic, two-year, inhalation exposure.

 \vspace{ 12 pt }
\linespread{1.2}
\begin{table} [htbp]
\begin{footnotesize}
\caption{Quantal carcinogenicity data:  Alveolar/bronchiolar adenomas and carcinomas in female B6C3F$_1$ mice after two-year inhalation exposure to cumene (C$_9$H$_{12}$).  Source: \cite{tr542}.}\label{tabl:data}
\renewcommand{\arraystretch}{1.25} 
\begin{center}
\begin{tabular}{l c c c c c}
\hline
Exposure conc.~(ppm), $d_i$  && 0 & 125 & 250 & 500  \\
\hline
Animals with tumors, $Y_i$ & & 4 & 31 & 42 & 46 \\
Animals tested, $N_i$ & & 50 & 50 & 50 & 50 \\
\hline
\end{tabular}
\end{center}
\end{footnotesize}
\end{table}

In the table, a clear dose response is evidenced. Of additional interest, however, is characterization of the effects at `low-dose' exposures, in order to inform risk characterization of this potential carcinogen.  We return to these data below.
 \vspace{6pt}

A contemporary approach to quantal-response risk estimation is known as \emph{benchmark analysis}. First introduced
by \cite{crum84}, the method uses an assumed functional specification for $R(d)$ to provide low-dose estimates for the risk.  When applied in environmental or public health settings, however, adjustments are made for background or spontaneous effects.  Since $R(0)$ represents the background response for all subjects in a population, the risk function is re-expressed in terms of an excess-above-background response rate \citep[\S4.2.1]{piba05}.  With quantal data, the \emph{extra risk function} $R_E(d)=\{R(d)-R(0)\}/\{1-R(0)\}$ is typically employed.  From this, the \emph{benchmark dose} (BMD) is calculated  by inverting $R_E(d)$ at a predetermined level of risk; the latter is called the \emph{benchmark risk} or \emph{benchmark response} (BMR).  In effect, we solve $R_E$(BMD) = BMR $\in (0,1)$; BMRs between 0.01 and 0.10 are most often seen in practice
\citep{epa12}.

Statistical inferences in benchmark analysis focus on $100(1-\alpha)\%$ confidence (or credible) limits for the BMD.  Driven by public health or other safety considerations, only one-sided, lower limits are targeted, denoted as BMDLs \citep{crum95}.  Where needed for clarity, we add a subscript for the BMR level at which each quantity is calculated:  BMD$_{\text{\tiny 100BMR}}$ and BMDL$_{\text{\tiny 100BMR}}$.
In this fashion, use of BMDs and BMDLs for quantifying and managing risk with a variety of endpoints is growing in both the United States and the European Union \citep{gao01,eu03,oecd06,oecd08}.

\subsection{\normalsize Parametric Bayesian Benchmark Analysis}\label{sec:Parametric BMD}

\indent\indent With quantal data such as those in Table \ref{tabl:data}, we take $Y_i$ as the number of responses at the $i^{\mbox{\scriptsize th}}$ dose level, out of $N_i$ subjects tested at that dose $(i = 1, ..., m)$.  The standard statistical model here is the binomial, i.e., $Y_i\sim \mbox{indep. Bin}\bigr(N_i, R(d_i)\bigl)$, where $R(d_i)$ is the risk at dose $d_i$. For generic purposes, we denote {\boldmath$\theta$} as an unknown parameter vector that describes $R(d)$.

In risk-analytic studies with quantal data, estimation of the BMD has traditionally been performed via maximum likelihood
\citep[\S4.3.1]{piba05}, producing maximum likelihood estimators (MLEs) $\widehat{\mbox{\boldmath $\theta$}}$, with corresponding MLEs $\widehat{R}(d)$ for $R(d)$ and $\widehat{R}_E(d)$ for $R_E(d)$.  Setting $\widehat{R}_E(d) =$ BMR and solving for $d$ yields the MLE, $\widehat{\mbox{BMD}}_{\text{\tiny 100BMR}}$.
The corresponding, frequentist BMDL is built from the statistical properties of $\widehat{\mbox{BMD}}$.

Appeal to Bayesian modeling and \emph{credible} limits for building BMDLs is far less developed.  For quantal-response data, \cite{nauf09} studied hierarchical dose-response models under a suite of forms available from the U.S. Environmental Protection Agency's Benchmark Dose Software, BMDS \citep{davi12}, including the well-known logistic and probit models: $R(d) = \{1 + \exp(-\beta_0 - \beta_1d)\}^{-1}$ and $R(d) = \Phi(\beta_0 + \beta_1d)$, respectively.  [Here, $\Phi(z)$ is the standard normal cumulative distribution function.]  They placed non-informative priors on the dose-response parameters of each model.  \cite{shsm11} also studied the logistic model, along with the quantal-linear model
\begin{equation}\label{eq:M3}
R(d) = 1 - \exp(-\beta_0 - \beta_1 d)
\end{equation}
(where $\beta_0 \ge 0$ and $\beta_1 \ge 0$), a popular dose-response form in risk assessment \citep{buck09}.  They similarly placed objective priors on the $\beta$-parameters; further developments appeared in \cite{shsm12}.  \cite{shao12} expanded these considerations to the probit model, and also introduced a power prior to build historical control information into the hierarchy. \cite{whba12} presented semi-parametric Bayesian models for the dose response, incorporating a probit kernel and cubic B-splines.  They placed normal priors on the basis-function coefficients and built dose-response monotonicity into their prior constraints. The normal priors were often flattened to be essentially non-informative (or \emph{objective}) although like \cite{shao12}, \citeauthor{whba12} built informative priors at
$d = 0$ to incorporate potential historical control information.  \citet{guro13} described a nonparametric Bayesian model for the quantal setting, with beta/Dirichlet priors on pertinent probabilities related to their nonparametric construction.

The Bayesian approach for BMD estimation has also been applied with \emph{non}-quantal data, and as in the quantal setting its use enjoys increasing application.  See, e.g., \cite{whba09b}, \cite{moib06}, \cite{held04}, or \cite{come03}.

Notably, previous methods for calculating parametric Bayesian BMDs typically assume objective (and sometimes improper) prior distributions for the unknown model parameters.
For example, in the logistic model mentioned above the $\beta$-parameters are interpreted as
the usual regression intercept and slope on a logit scale.  With quantal data, at least, it is unusual for truly informative, risk-analytic, prior knowledge to be available on parameters  such as $\beta_0$ and $\beta_1$, since their interpretation is so generic.
Such a strategy
neglects informative prior information that may be available on the true quantity of interest in this setting, the BMD.

Of course, informative prior knowledge on the BMD may not always exist in practice, and in this case objective priors serve a useful purpose.  When such prior information is available, however, the Bayesian paradigm can achieve its true potential.  Herein, we explore this by reparameterizing the quantal-response model to explicitly incorporate the BMD.
Following on previous presentations for parametric Bayesian benchmark modeling \citep{shsm11, shsm12}, we focus on the popular quantal-linear dose-response model in \eqref{eq:M3}, due to its heavy use in environmental toxicology.
Section \ref{sec:modelM3} describes our reparameterization strategy in greater detail.
Section \ref{sec:BBA} develops a Bayesian framework using the reparameterized quantal-linear form
for $R(d)$. Section \ref{sec:exM3} returns to the cumene carcinogenicity data from Example 1,
while Section \ref{sec:disc} ends with a brief discussion.

\section{The Quantal-Linear Dose-Response Model} \label{sec:modelM3}
\subsection{Benchmark Analysis with the Quantal-Linear Model} \label{sec:modelM3intro}

\indent\indent The quantal-linear form $R(d) =  1 - e^{-\beta_0 - \beta_1d}$ from \eqref{eq:M3} is a popular
construct for dose-response modeling.  Also known as the one-stage model or the complementary-log model, it is a special case of the famous multistage model in carcinogenicity testing \citep[\S4.2.1]{piba05}.  This has led it to wide application in toxicological risk assessment with quantal data \citep{foro07,buck09,shsm12}.
Notice that $R(0) = 1 - e^{-\beta_0}$ so that the quantal-linear extra risk is simply 
$R_E(d)=1-e^{-\beta_1d}$.
To find the BMD, we solve $R_E(\mbox{BMD})=\mbox{BMR}$ to obtain
\begin{equation}\label{eq:M3bmd}
\mbox{BMD}_{\text{\tiny 100BMR}} = -\frac{\log(1-\mbox{BMR})}{\beta_1}.
\end{equation}
Recall that we require $\beta_j \ge 0 \ (j = 0,1)$ in \eqref{eq:M3}.
[In the exceptional setting where $\beta_1 = 0$,
no dose effect is present.  If so, the agent under study would be viewed as innocuous, consistent with the corresponding `definition' of BMD $= \infty$ under \eqref{eq:M3bmd}.]

\subsection{Reparameterizing the Quantal Linear Model}\label{sec:Reparameterization}


\indent\indent For benchmark risk analysis, a risk assessor, toxicologist, or other domain expert may often have prior knowledge about the target parameter, the BMD, and possibly also about other application-specific values such as the risk at certain doses; cf.~the use of control-response information for $R(0)$ by \cite{shao12} and \cite{whba12}.
Our goal is to
reparameterize $R(d)$ in terms of meaningful parameters whose prior distributions are more intuitive to elicit in practice.
The reparameterization strategy is, of course, not new, even in benchmark analysis; e.g., \citet[\S14.3.4]{papo05} re-expressed the quantal-linear model in terms of the BMD to facilitate construction of frequentist BMDLs.  Following on their lead, we reformulate the unknown regression parameters in \eqref{eq:M3} in terms of well-understood risk-analytic quantities: the target value, BMD, and the background risk, $R(0)$.
[The latter quantity is technically a nuisance parameter here, but it is nonetheless likely to be associated with non-trivial prior information, e.g., from historical control databases \citep{whba12}.] For simplicity, denote these as $\xi$ = BMD and $\gamma_0 = R(0)$, respectively.  Under
\eqref{eq:M3}, we know $\xi = -\log(1-\mbox{BMR})/\beta_1$ and $\gamma_0 = 1 - e^{-\beta_0}$.
%
%
%
%
[For convenience in the sequel, we generally suppress the BMR subscript on $\xi$, although it is understood that the BMD is dependent on the chosen level of BMR $\in (0,1)$.]
With these, $R(d)$ becomes:
\begin{eqnarray}\label{eq:M3R}
R(d)=1-\exp\left\{\log(1-\gamma_0)+\frac{\log(1-\mbox{BMR})}{\xi}d\right\}.
\end{eqnarray}
The corresponding extra risk is simply $R_E(d) = 1 - (1-\mbox{BMR})^{d/\xi}$.
While the notation may appear more burdensome, explicit incorporation of the parameters $\xi$ and $\gamma_0$ allows us to formulate a clearer, more application-oriented hierarchical model.  We explore this further in the next section.

\section{Bayesian Benchmark Analysis}\label{sec:BBA}
\subsection{Prior Specification}\label{sec:prior}
\indent\indent Under our reformulation the unknown parameter vector is ${\boldsymbol \theta} = [\xi \ \gamma_0]^T$.
To construct a Bayesian hierarchy, we assign a joint p.d.f.~to $\mbox{\boldmath $\theta$}$: $\pi(\mbox{\boldmath $\theta$}) = \pi(\xi, \gamma_0)$.  Mimicking previous Bayesian models for benchmark analysis \citep{shsm11,shao12}, we assume the unknown parameters enter into the prior independently, so that $\pi(\xi, \gamma_0) = \pi(\xi)\pi(\gamma_0)$.  This assumption is not as arbitrary as it may seem:  to assess potential correlation between the two quantities, we accessed data from a study \citep{nitc07} of 91 different chemical carcinogens archived in the U.S.~EPA's Integrated Risk Information System (IRIS: see \verb|http://www.epa.gov/iris|).  Each of the 91 IRIS data sets represented a quantal dose-response to a particular carcinogen, allowing us to compute (frequentist) ML estimates for both $\xi$ and $\gamma_0$ under
\eqref{eq:M3}.  We then calculated the Spearman rank correlation $r_S$ \citep[\S7.3]{kvvi07} across the 91 pairs of MLEs $(\widehat{\xi} , \widehat{\gamma_0})$.  We found the correlation to be insignificantly different from zero: $r_S = -0.039; P > 0.20$.

To specify the individual components of the joint prior, we employ established, flexible forms.  For $\xi$ we make an inverse gamma assumption: $\xi \sim \mbox{\textit{IG}}(\alpha, \beta)$ with marginal prior density $\pi(\xi|\alpha,\beta) = \frac{\beta^\alpha}{\Gamma(\alpha)}\xi^{-(\alpha+1)}e^{-\beta/\xi}I_{(0,\infty)}(\xi)$, where $\Gamma(a)$ is the usual gamma function and $I_{\mathbb{A}}(x)$ is the indicator function that returns 1 if $x \in \mathbb{A}$ and 0 otherwise.  For $\gamma_0 = R(0)$ we take $\gamma_0 \sim \mbox{\textit{Beta}}(\psi, \omega)$ with marginal prior $\pi(\gamma_0|\psi,\omega) = \frac{\Gamma(\psi+\omega)}{\Gamma(\psi)\Gamma(\omega)}\gamma_0^{\psi-1}(1-\gamma_0)^{\omega-1}I_{(0,1)}(\gamma_0)$. (Below, we examine how deviations from these priors---emphasizing sensitivity to the inverse gamma assumption---affect the eventual inferences on $\xi$.)
Conceptually, this approach shares similarities with a Bayesian inverse dose estimation strategy suggested by \citet{baro98} for use in clinical trials; however, technical differences exist between the two methods, driven primarily by dissimilarities between their medical-safety application and our environmental-risk motivation.

The various hyperparameters, $\alpha$, $\beta$, $\psi$, and $\omega$, require specification for implementation of our model.  (One could also build further levels into the hierarchy by constructing hyperprior p.d.f.s, and this has potential for future study.  We do not investigate it here, however.)  Our goal is elicitation, to the best extent possible, of each marginal prior by incorporating the domain expert's prior knowledge.

A broad literature exists on how to conduct prior elicitation; see, e.g., \cite{ohbu06} or \cite{kuhn11} and the references therein.  From it, we adopt the general strategy that domain experts with minimal statistical expertise are best able to provide prior information in the form of basic location summaries.
Medians and other quartiles can be especially effective:  based on interactions with environmental toxicologists and risk assessment domain experts, we have found that for the target parameter $\xi$, specification of the the median ($Q_2$), along with the first/lower quartile ($Q_1$) of the IG prior is most effective.  Put simply, prior expert knowledge for $\xi$ is likely to be more accurate closer to the origin, since BMDs are associated with adverse effects at low doses.  Similarly, for the beta prior on $\gamma_0$ we also elicit the two quartiles $Q_1$ and $Q_2$. We then solve for the pertinent hyper-parameters given these quartiles.  (Details are provided in a Supplemental Document.)  Of course, other elicitation values are possible, and analysts may wish to experiment with selection of other quartiles, terciles, percentiles, etc.

Given the resulting values of $\alpha$, $\beta$, $\psi$, and $\omega$, the joint prior for ${\boldsymbol \theta} = [\xi \ \gamma_0]^T$ is fully specified, and we proceed with calculation of the posterior p.d.f.
We describe this in the next subsection. Before continuing, however, we acknowledge that cases can arise where the prior elicitation breaks down, say, if prior experience with an environmental agent
is so limited that the toxicologist simply has no idea where the BMD will lie. When this occurs, moving to objective specifications for the prior densities may be necessary.  Many possibilities exist for building objective priors on a strictly positive quantity such as $\xi$, and on a probability such as $\gamma_0$. We favor a simple approach: for an objective prior on $\xi$, use $\xi \sim {IG}(0.001,0.001)$ [another option could be $\xi \sim {Gamma}(0.001,0.001)$; see below].  The IG prior is a popular suggestion in the literature for right-skewed, positive quantities \citep[\S1.2]{lamb05,chjo11}, such as the BMD. One could alternatively appeal to an (improper) objective prior of the form $\pi(\xi) \propto 1/\xi$, a conventional choice that is often recommended for positive parameters \citep[\S9.17]{ohag94}.  In order to facilitate the posterior computations, however, we prefer to employ proper prior densities. In fact, the ${IG}(0.001,0.001)$ prior can approximate the improper reciprocal prior quite well when $\xi$ is not extremely close to zero. The ${Gamma}(0.001,0.001)$ can also serve as an approximator for $1/\xi$, although for large dose values the approximation can break down.  Indeed, we find that scaling the doses to make the highest dose equal 1 is a convenient device that allows both ${IG}(0.001, 0.001)$ and ${Gamma}(0.001,0.001)$ to approximate the conventional improper $1/\xi$ prior. As such, we perform our calculations with doses scaled so that the maximum administered dose equals 1.

For an objective prior on $\gamma_0$, we consider the univariate Jeffreys prior:  $\gamma_0 \sim \mbox{\textit{Beta}}\bigl(\frac12,\frac12\bigr)$.  This assigns symmetrically high density to values of $\gamma_0 \to 0$ and $\gamma_0 \to 1$.  The former is reasonable in our toxicological setting, although the latter is problematic, since $\gamma_0$ represents the probability of an adverse response under no exposure and is usually small in risk-analytic applications.  (A similar complaint could be raised against use of a `vague' uniform prior for $\gamma_0$.) In the end, even the valid-but-imprecise argument that low prior weight should be assigned to values of $\gamma_0 \rightarrow 1$ is itself a form of subjective prior knowledge, and construction of truly objective priors for $\gamma_0$ may be difficult.  As the background response rate in the subject organism under study, $\gamma_0$ will be a well-understood quantity in practice, and as others have noted there will often be at least some useful historical control information available for it.  We expect that calls for objective priors on $\gamma_0$ in this setting will be rare. If in the extreme this is not the case, we default to the objective Jeffreys prior $\gamma_0 \sim \mbox{\textit{Beta}}\bigl(\frac12,\frac12\bigr)$, although we recognize the enigmatic aspects of such a strategy.

\subsection{Posterior Analysis via Stochastic Approximation}\label{sec:posterior}
\indent \indent Given our prior specification from \S\ref{sec:prior} under a binomial likelihood,
the joint posterior p.d.f.~for ${\boldsymbol \theta} = [\xi \ \gamma_0]^T$ is
\begin{equation}\label{eq:post2}
\pi(\xi, \gamma_0|{\boldsymbol Y})=\frac{\prod_{i=1}^n{N_i \choose Y_i}{R(d_i)}^{Y_i}{\{1-R(d_i)\}}^{N_i-Y_i}}{m(\boldsymbol Y)}
\frac{\beta^\alpha e^{-\beta/\xi}}{\Gamma(\alpha)\xi^{\alpha+1}}
\frac{\Gamma(\psi+\omega)}{\Gamma(\psi)\Gamma(\omega)}\gamma_0^{\psi-1}(1-\gamma_0)^{\omega-1}
\end{equation}
over $\xi > 0$ and $\gamma_0 > 0$.  The denominator of $\pi(\xi, \gamma_0|{\boldsymbol Y})$ contains the marginal likelihood
\[m(\boldsymbol Y) = \int_0^\infty\int_0^1 \prod_{i=1}^n{N_i \choose Y_i}{R(d_i)}^{Y_i}{\{1-R(d_i)\}}^{N_i-Y_i}\pi(\xi|\alpha,\beta)\pi(\gamma_0|\psi,\omega)d\gamma_0d\xi,\]
which is intractable under our elicited prior structure. To evaluate \eqref{eq:post2} we therefore turn to Monte Carlo posterior approximations using an adaptive Metropolis (AM) strategy \citep{anth08}.  This is detailed in the Supplementary Document.

For a given data ensemble $(\boldsymbol Y, \boldsymbol N, \boldsymbol d)$, we employ the AM algorithm to produce a bivariate chain of draws, $\{\xi_k,\gamma_{0k}\}_{k=1}^K$. For practical purposes, however, we first check the acceptability of the data before employing the algorithm. Our experience shows that a shallow dose response can create unstable frequentist $\hat{\xi}_{\mbox{\tiny 100BMR}}$ estimates, and very shallow responses may cause the model fit to fail entirely.
Indeed, the EPA's BMDS software program will not estimate a BMD for a flat or negatively trending dose response \citep{whba09a}. For these reasons, we chose to mark flat or decreasing-trend data as `data failures' and not perform estimation for them (see the Appendix).

We apply the AM algorithm to any data set passing our data-failure screen.
We also include a `burn-in' diagnostic---detailed in the Supplementary Document---to ensure that the Monte Carlo chain produces a stable approximation to the posterior distribution.  If the diagnostic fails after multiple generations of the AM chain, we consider this an `algorithm failure' and do not report values for the BMD or BMDL.  (This is a rare occurrence.  Again, see the Supplementary Document for specifics.)
The larger the value of $K$ and the more stable the resulting chain, as determined by the diagnostics mentioned above, the better the approximation. Our experience suggests that a starting value of $K = 100,000$,
with reduction for burn-in to about 90,000 or 80,000 draws,
generally provides stable results.  We derive inferences on $\xi$ from this retained sample of draws.

\subsection{Bayesian Estimation and Inference on the BMD} \label{sec:estimation}

\indent \indent For estimating the BMD, we appeal to standard decision-theoretic principles and select the Bayes estimator based on minimization of the Bayes risk, after specification of an underlying loss function.  As is well-known, choice of squared-error loss leads to the posterior mean,
$\text{E}[\xi|\boldsymbol Y]$, as the Bayes estimator, while absolute-error loss leads to the posterior median \citep[\S7.3.4]{cabe02}.
We can approximate these posterior quantities using our Monte Carlo sample's arithmetic mean or median, respectively.
If subject-matter considerations cannot guide the choice of loss function, and thus which estimator to employ, appeal to absolute-error loss and the posterior median might be preferable.  This suggestion is, admittedly, empirical: with small sample sizes we generally find the (approximated) posterior for $\xi$ to exhibit a right skew, and the median is more robust against large skews.
This has an important, practical consequence:  note that a larger estimated BMD implies a higher level of `acceptable' exposure to a potential toxic agent.  If this estimated value is artificially inflated due to anomalies such as a heavy right skew, any public health or environmental guidelines based on the estimate may be unnecessarily lax, and even unsafe. Using the posterior median rather than the posterior mean in such a situation would represent a more-precautionary course of action.

Alert readers may notice that from a decision-theoretic perspective, this issue of how to estimate the BMD begs a larger question: both the squared-error and absolute-error losses treat deviations symmetrically. Should we employ an \emph{asymmetric} loss function here instead?
Arguably, yes. In effect, a high BMD quantitatively views large exposures to a potentially hazardous agent as relatively safe.  If, in truth, the agent is highly toxic, the consequences of such a decision could be severe from a public health or environmental safety perspective.  This is more consequential than incorrectly driving BMD $\rightarrow 0$ and imposing strict exposure limitations on an innocuous or weakly toxic agent. Unless (considerable, we would contend) socioeconomic factors can counterbalance these safety concerns, BMD overestimation generates a greater `loss'.

Many asymmetric loss constructions are possible; we employ the simple bilinear loss function of \citet[\S2.46]{ohag94}:
\begin{equation}\label{eq:bilin}
L(\Delta,\xi) = \left\{\begin{array}{cc}
a(\xi-\Delta)&\Delta\le{\xi}\\
b(\Delta-\xi)&\Delta>{\xi}\\
\end{array} \right.
\end{equation}
where $\xi$ is the target quantity, estimated by the decision $\Delta$. The constants $a$ and $b$ tune the bilinear loss for each individual application.
(When $a = b$ we recover absolute-error loss.) Since we treat overestimation of $\xi$ more harshly than underestimation, we take $b > a > 0$. \cite{ohag94} shows that under \eqref{eq:bilin}, the optimal Bayes estimator for $\xi$ is the $100\left(\frac{a}{a+b}\right)$th percentile of the posterior distribution $\pi(\xi|{\boldsymbol Y})$.
As a default choice for benchmark dose estimation, we suggest setting the ratio $\frac{a}{b}=\frac{1}{2}$, i.e., overestimation of $\xi$ incurs twice as much relative loss as underestimation.  If so, the optimal Bayesian estimator $\widehat{\xi}_{\text{\tiny 100BMR}}$ becomes the lower/first tercile of $\pi(\xi|{\boldsymbol Y})$.  We estimate this with the lower tercile from our Monte Carlo sample of $\xi$.

For a Bayesian BMDL, say, \underline{$\xi$}$_{\text{\tiny 100BMR}}$, we essentially desire a one-sided, lower, $100(1-\alpha)$\% credible limit on $\xi$, satisfying $P(\xi >$ \underline{$\xi$}$_{\text{\tiny 100BMR}}|\boldsymbol Y)=1 - \alpha$.  At the traditional level of $\alpha = 0.05$, this is the lower 5$^{\scriptsize \mbox{th}}$ percentile of $\pi(\xi|{\boldsymbol Y})$, which we approximate via the lower $5^{\scriptsize \mbox{th}}$ percentile from our Monte Carlo chain.

\subsection{Prior Sensitivity: $\boldsymbol \epsilon$-Contamination Analysis}\label{sec:epsilon contam}
\indent \indent To investigate the influence of our IG prior assumption on the eventual BMDL \underline{$\xi$}$_{\text{\tiny 100BMR}}$, we consider a prior sensitivity analysis.  (One could also perform a sensitivity analysis of the beta prior for $\gamma_0$; however, since $\gamma_0$ is an obvious nuisance parameter in terms of BMD estimation and inference, we do not highlight that alternative here.)  Our experiences with $\xi$ for this model show that the parameter uncertainty can almost always be described via a right-skew, which helps motivate the IG prior specification.  As mentioned above, however, an obvious alternative is the traditional gamma distribution: $\xi \sim \mbox{\textit{Gamma}}(\alpha, \beta)$, where $\alpha$ and $\beta$ are elicited using the same expert opinion as detailed above.

We follow \citet[\S7.15]{ohag94} and rewrite the prior density on $\xi$ as a contaminated mixture of IG and gamma priors. Treating the IG density as the base prior, $\pi_0(\xi)$, and the gamma as the contaminating density, $q(\xi)$, the prior for $\xi$ is written as
\begin{equation}\label{eq:epscon}
\pi(\xi)=(1-\epsilon)\pi_0(\xi)+\epsilon q(\xi),
\end{equation}
where $\epsilon\in[0,1]$ controls the degree of prior contamination. When $\epsilon=0$, no contamination appears and the prior for $\xi$ is the base IG density; when $\epsilon=1$, the prior density function for $\xi$ is completely replaced by the contaminating gamma prior.

Employing the prior density in the $\epsilon$-contamination form from \eqref{eq:epscon}, we denote \underline{$\xi$}$_{\text{\tiny 100BMR}}(\epsilon)$ as the estimated BMDL for a specific $\epsilon$. We then monitor the evolution of \underline{$\xi$}$_{\text{\tiny 100BMR}}(\epsilon)$ as $\epsilon$ increases from 0 to 1.

An additional quantification of the $\epsilon$-contaminating prior's influence is the instantaneous rate of change of the posterior inference at $\epsilon=0$  \citep[\S7.15]{ohag94}. We calculate this as the (absolute value of the) first derivative of \underline{$\xi$}$_{\text{\tiny 100BMR}}(\epsilon)$ at $\epsilon=0$: $|D(q)|=\left|\lim_{\epsilon\rightarrow0}\frac{1}{\epsilon}\{\mbox{\underline{$\xi$}$_{\text{\tiny 100BMR}}(\epsilon)$}-\mbox{\underline{$\xi$}$_{\text{\tiny 100BMR}}(0)$}\}\right|$.
%
%
\noindent \citeauthor{ohag94} showed that $D(q)$ simplifies to
\begin{equation} \label{eq:Dq}
D(q) = \left\{\mbox{\underline{$\xi$}$_{\text{\tiny 100BMR}}(1)$}-\mbox{\underline{$\xi$}$_{\text{\tiny 100BMR}}(0)$}\right\}\frac{m_q(\boldsymbol Y)}{m_0(\boldsymbol Y)},
\end{equation}
where $m_q(\boldsymbol Y)$ and $m_0(\boldsymbol Y)$ are the marginal likelihoods under the contaminating gamma prior density, $q(\xi)$, and base IG prior density, $\pi_0(\xi)$, respectively.

In order to calculate $|D(q)|$, values of \underline{$\xi$}$_{\text{\tiny 100BMR}}(1)$ and \underline{$\xi$}$_{\text{\tiny 100BMR}}(0)$ are obtained as the estimated BMDLs using $q(\xi)$ and $\pi_0(\xi)$, respectively. To find the marginal likelihoods, $m_q(\boldsymbol Y)$ and $m_0(\boldsymbol Y)$, we appeal to our AM sample and employ a geometric estimator via the bridge sampling method recommended by \cite{meng96} and \cite{lope04}. (Details are given in the Supplementary Document.)
$|D(q)|$ can then be used as a measure of sensitivity:
as it draws closer to zero, the posterior inferences are less affected by the contaminating prior \citep[\S7.15]{ohag94}.
We apply $|D(q)|$ as part of our larger Bayesian benchmark analysis.  The next section illustrates these various calculations.

\subsection{Posterior Calculations on the Extra Risk Scale}\label{sec:ExtraRisk}
\indent\indent Despite its moniker, the benchmark dose is critically related to the extra risk function $R_E(d)$.  As \citet{wepi12} emphasized, the BMD and in particular the BMDL are used to indicate where exposures to hazardous stimuli lead to extra risks at or below the targeted BMR, since they serve as points of departure for an environmental risk assessment.  The construction of a joint posterior for $\xi$ and $\gamma_0$ allows for convenient exploration of features on the extra risk scale, by querying their corresponding draws from the bivariate AM-chain.

\subsubsection{Posterior extra risks at selected doses}\label{sec:ExtraRiskPosterior}
\indent\indent Under our reparameterization,
a one-to-one correspondence will typically exist between the pair $(\gamma_0,\xi)$ and the $R_E(d)$ curve in the extra risk space. For example, with the quantal-linear model the extra risk $R_E(d) = 1 - (1-\text{BMR})^{d/\xi}$ is clearly dependent upon $\xi$ (and is independent of $\gamma_0$, although this is not guaranteed in general): as $\xi$ increases, the quantal-linear $R_E(d)$ curve becomes less concave and drops towards zero in a consistent fashion.
Such correspondences allows us to explore the posterior distribution of the extra risk at any given $d$, after obtaining the joint posterior distribution of ($\gamma_0, \xi$).

For example, a risk assessor may be interested in the posterior distribution of the extra risk at certain crucial or regulatory dose levels.  Given a posterior AM sample of ($\gamma_0, \xi$) pairs, sans burn-in, one can map these to the corresponding posterior sample of $R_E(d)$ curves.  Then, the vertical intersection of this sample of curves at various dose levels $d$ corresponds to a posterior sample of the extra risk at that $d$. Density estimates of this posterior can be constructed via standard kernel methods, allowing the analyst to compare the posterior densities across a variety of pertinent doses; the latter could include the BMDL, the BMD estimate (or estimates, if different selection criteria are under consideration), etc.  We explore this possibility with the cumene data in \S\ref{sec:pstER}, below.

\subsubsection{Simultaneous credible bands}\label{sec:ExtraRiskBands}
\indent\indent This correspondence between $(\gamma_0,\xi)$ pairs and $R_E(d)$ curves open up a panoply of inferences on the extra risk scale.  For instance, a posterior `point' estimate of $R_E(d)$ corresponds to a point estimate of $(\gamma_0,\xi)$.  The simplest example would be calculation of the mean centroid $(\bar{\xi},\bar{\gamma}_0)$ from the retained AM chain, since it is a well-defined quantity.
(Bivariate medians are not always unambiguously defined. And, the sort of asymmetric interest described above for estimating the BMD is not as crucial for estimating the extra risk.  Thus a `central' value such as a mean-based centroid would be a reasonable default.)  The curve in $R_E$-space corresponding to $(\bar{\xi},\bar{\gamma}_0)$ serves as an estimator, $\hat{R}_E(d)$, for the extra risk.

Further, a $1-\alpha$ \emph{simultaneous credible band} can be constructed in a straightforward fashion.  Simply identify a central subset of the retained AM chain containing $100(1-\alpha)$\% of the $(\gamma_0,\xi)$ pairs and sweep out the envelope in $R_E$-space to which these points correspond.

Risk-analytic operations generally focus attention on strictly upper limits for the extra risk---corresponding to lower limits on the BMD.  This leads to one-sided, upper, credible bands, and simplifies the band construction.  Simply find the `smallest' retained $(\gamma_0,\xi)$ AM-pair above which $100(1-\alpha)$\% of the AM-draws lie.  The corresponding $R_E(d)$ curve will serve as an upper band on the extra risk, valid simultaneously over all $d \ge 0$.  How to define `smallest' is, of course, open to consideration.  As an initial proposal, we suggest recognizing $\gamma_0$'s nuisance status and focusing on how $\xi$ varies throughout $(\gamma_0,\xi)$-space: find the smallest value of $\xi$ above which $100(1-\alpha)$\% of the $(\gamma_0,\xi)$ AM-pairs lie.  [One can view this as a horizontal supporting line---or a supporting hyperplane for models with more than two parameters---separating the upper $100(1-\alpha)$\% of the retained chain from the lower $100\alpha$\% in the $(\gamma_0,\xi)$ plane.]  In effect, this reduces attention to the marginal posterior for $\xi$ and isolates the $(\gamma_0,\xi)$ pair at $\xi$ = \underline{$\xi$}$_{\text{\tiny 100BMR}}$.  The corresponding curve in $R_E$-space serves as the $1-\alpha$ upper band.

\section{Cumene Carcinogenicity Data, Revisited}\label{sec:exM3}
\subsection{Benchmark data analysis}\label{sec:exM3 result}
\indent\indent We applied our Bayesian approach with the reparameterized quantal linear model to the cumene carcinogenicity data in Example \ref{sec:intro}.
Notice that the C$_9$H$_{12}$ exposure dose, $d$, is actually a concentration (in ppm) here, and so technically we will compute benchmark concentrations (BMCs) based on the quantal carcinogenicity data.
We operated with BMR set to the traditional level of 0.10 \citep{epa12}. Using input from domain experts, the prior elicitation was based on existing background from the toxicological literature.  Table \ref{tabl:elicitation} summarizes the elicited values. (Details are given in the Supplemental Document.)  The consequent prior distributions employed in the analysis were $\xi\sim IG(0.53,0.13)$ and $\gamma_0\sim Beta(1.36,12.31)$.  We then applied our AM approach to approximate the posterior for $\xi$ and $\gamma_0$. No `data failures' or `algorithm failures' were encountered.

  \vspace{ 12 pt }
\linespread{1.2}
\begin{table} [htbp]
\begin{footnotesize}
\caption{Prior elicitation summary for cumene carcinogenicity data.}\label{tabl:elicitation}
\renewcommand{\arraystretch}{1.25} 
\begin{center}
\begin{tabular}{c c c c c c l c}
\hline
Parameter  & & Lower quartile, $Q_1$ & Median, $Q_2$  & &  \multicolumn{3}{c}{Prior parameters}\\
\hline
$\xi$      & &  90 ppm (0.18*)       & 250 ppm (0.50*) & & $\alpha =$ 0.53 &  & $\beta =$ 0.13\\
$\gamma_0$ & &  0.04                 & 0.08            & & $\psi =$ 1.86   &  & $\omega =$ 12.31\\
\hline
\multicolumn{6}{l}{\footnotesize * after scaling to unit interval}
\end{tabular}
\end{center}
\end{footnotesize}
\end{table}

We instituted an AM chain size of $100,000$, from which our convergence diagnostic procedure recommended an initial burn-in of $10,000$ draws.  The remaining $90,000$ draws were then employed as the posterior approximation for these data.  Trace plots showed adequate mixing of the chain; see the Supplemental Document.
Figure \ref{fig:f2} displays the corresponding histogram (with overlayed kernel density estimate) for the $\xi$ component. A unimodal and slightly right-skewed marginal posterior distribution for $\xi$ is indicated.

\begin{figure}[htbp]
\centering
\includegraphics[scale=0.3]{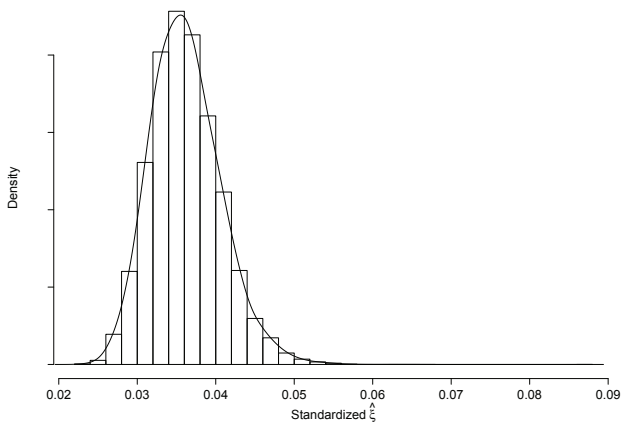}
\caption{\small Histogram and kernel density estimator for posterior approximation to $\xi_{10}$ in Example 1. Dose scale ($\xi$) is standardized to unit length after dividing by the highest dose in the data set.}\label{fig:f2}
\end{figure}

\begin{figure}[htbp]
\centering
\includegraphics[scale=0.3]{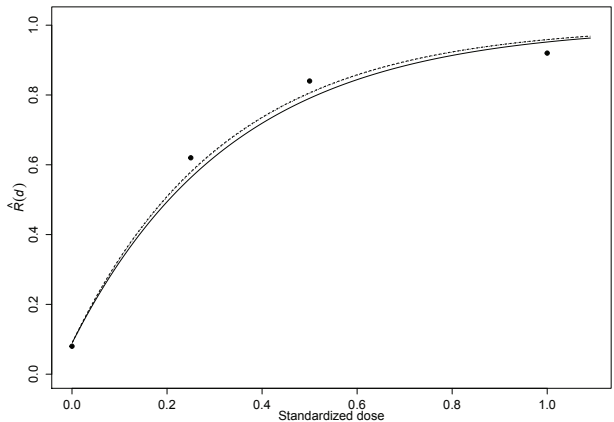}
\caption{\small Estimated risk functions for cumene carcinogenicity data in Example 1. Solid curve (\solidrule) is based on the posterior median, dashed curve (\protect\dashedrule) is based on the lower posterior tercile, and dotted curve
($\cdots\cdots$) is from the maximum likelihood estimate (MLE). The estimated risk function based on the lower tercile is indistinguishable from the estimated ML risk function at this scale. Solid circles are observed proportions. Dose scale ($\xi$) is standardized to unit length after dividing by the highest dose in the data set.}\label{fig:f3}
\end{figure}

Figure \ref{fig:f3} displays the estimated risk function $\hat{R}(d)$ based on the median posterior estimates for $(\xi, \gamma_0)$, the lower tercile estimates for $(\xi, \gamma_0)$, and the MLEs $(\hat{\xi}$, $\hat{\gamma}_0)$, along with the original observed proportions. All three curves give reasonable estimates for $R(d)$, relative to the data, although the median-based Bayesian estimate shifts consistently to the right of the other two risk functions at most levels of $d$.

For reporting purposes, the final benchmark estimates (when translated back to the original ppm scale) are $\hat{\xi}_{10}$=17.973 ppm if using the sample posterior median or $\hat{\xi}_{10}$=17.046 ppm if using the sample posterior lower tercile.
The 95\% BMCL is \underline{$\xi$}$_{10}= 14.752$ ppm.
Comparing these to standard frequentist estimates, the MLE is $\hat{\xi}_{10}$=17.062 ppm and a 95\% frequentist Wald BMCL is \underline{$\xi$}$_{10}= 13.618$ ppm.
Both sets of values rest in similar ranges, and provide comparable points of departure for conducting further risk-analytic calculations on cumene carcinogenicity.

Thus for these data, our Bayesian approach operates similarly to the frequentist analysis, but it also provides an additional benefit for risk assessors: the Bayesian strategy combines elicited prior information on both the background risk and the BMC, potentially improving the estimation process by incorporating more-complete prior information.


\subsection{Prior sensitivity}\label{sec:prior sensitivity}
\indent\indent To explore sensitivity of the IG prior for these data, we returned to the methods in \S\ref{sec:epsilon contam} and contaminated the base prior for $\xi$ via three scenarios. In Scenario 1, we chose an objective $IG(0.001, 0.001)$ as the base prior for $\xi$ and contaminated it by the similar, objective, $Gamma(0.001,$ $ 0.001)$ prior.
This scenario was used to investigate the sensitivity of the $IG(0.001$, $0.001)$ prior for \underline{$\xi$}$_{\mbox{\tiny 100BMR}}$ when no elicitation is available.
In Scenario 2, we chose the elicited IG prior as the base prior for $\xi$ and contaminated it by a similarly elicited gamma prior. We used the same quartile information for $\xi$ to build both elicited priors. Scenario 2 was used to investigate the sensitivity of the IG prior
when elicitation is available.
In Scenario 3, we chose the elicited IG prior as the base prior for $\xi$ and contaminated it by an objective $Gamma(0.001,0.001)$ prior. Scenario 3 was used to investigate the robustness of \underline{$\xi$}$_{\mbox{\tiny 100BMR}}$ when the elicited IG prior is contaminated by an objective gamma prior.
Throughout, the prior for $\gamma_0$ was taken as either the elicited beta prior, $Beta(1.36,12.31)$, or an objective $Beta\big(\frac12,\frac12\big)$ prior. This gave six different settings for study, under which we monitored the consequent BMDL \underline{$\xi$}$_{\mbox{\tiny 100BMR}}(\epsilon)$ as $\epsilon$ increased from 0 to 1. The corresponding $|D(q)|$ instantaneous change measures from \eqref{eq:Dq} were also calculated.

\begin{figure}[!htbp]
\centering
\includegraphics[scale=0.35]{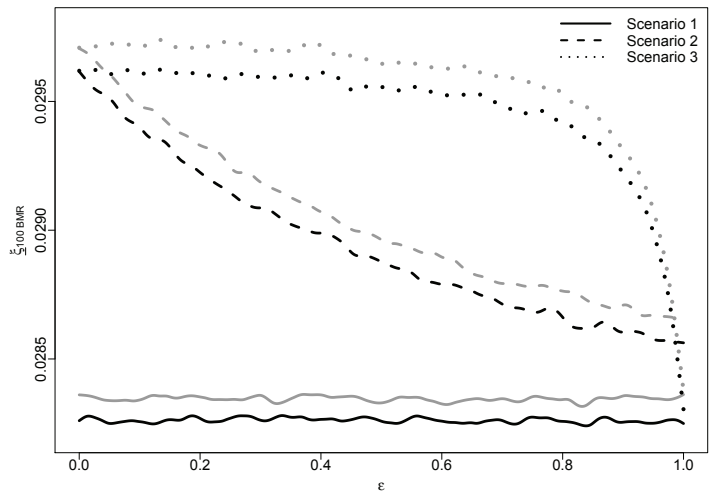}
\caption{\small Kernel smoothed value of \underline{$\xi$}$_{\mbox{\tiny 10}}$ as a function of changing $\epsilon$ in $\epsilon$-contamination study with cumene carcinogenicity data from Table \ref{tabl:data}. Solid curves (\solidrule) are from Scenario 1: an objective $IG(0.001,0.001)$ prior for $\xi$ is contaminated by an objective $Gamma(0.001,0.001)$ prior. Dashed curves (\dashedrule) are from Scenario 2: the elicited IG prior for $\xi$ is contaminated by the elicited gamma prior. Dotted curves
($\cdots\cdots$)
are from Scenario 3: the elicited IG prior for $\xi$ is contaminated by an objective $Gamma(0.001,0.001)$ prior. Gray curves indicate an objective $Beta\big(\frac12,\frac12\big)$ prior for $\gamma_0$; black curves indicate the elicited beta prior for $\gamma_0$. Dose scale ($\xi$) is standardized to unit length after dividing by the highest dose in the data set.}\label{fig:f4}
\end{figure}

Figure \ref{fig:f4} displays the evolution of \underline{$\xi$}$_{\mbox{\tiny 100BMR}}(\epsilon)$ across these six settings. The three choices for $\xi$ are distinguished using different line types (see the figure legend), and the two beta priors are distinguished using grey or black shading. In Scenario 1, no substantive change in \underline{$\xi$}$_{\mbox{\tiny 100BMR}}(\epsilon)$ is evidenced as $\epsilon$ varies from 0 to 1 in Figure \ref{fig:f4}. This suggests that the objective $IG(0.001,0.001)$ prior is essentially equivalent to an alternative objective $Gamma(0.001,0.001)$ prior in approximating the improper reciprocal prior, at least for these data.  For either prior on $\gamma_0$, the various \underline{$\xi$}$_{\mbox{\tiny 100BMR}}(\epsilon)$ values are also typically smaller than those from the other scenarios: the objective priors appear to consistently shrink the BMDL towards zero.

In Scenario 2, \underline{$\xi$}$_{\mbox{\tiny 100BMR}}(\epsilon)$ decreases in an exponentially decaying fashion, to a maximum drop of about $3-4\%$ as $\epsilon$ increases from 0 to 1 (Figure \ref{fig:f4}). This smooth decrease suggests that the contaminating gamma prior affects \underline{$\xi$}$_{\mbox{\tiny 100BMR}}$ in a consistent fashion as $\epsilon$ changes. However, the relatively small maximum decrease also suggests that the sensitivity of \underline{$\xi$}$_{\mbox{\tiny 100BMR}}$ to the contaminating elicited gamma is limited.

In Scenario 3, small decreases in \underline{$\xi$}$_{\mbox{\tiny 100BMR}}(\epsilon)$ are evidenced as $\epsilon$ increases from 0 to approximately 0.8, followed by a precipitous drop as $\epsilon$ increases to 1 (Figure \ref{fig:f4}). The maximum decrease is slightly more than $4\%$, however. This provides some evidence for the robustness of \underline{$\xi$}$_{\mbox{\tiny 100BMR}}$ under an elicited IG prior; a significant amount of contamination from the objective prior is required to greatly reduce the BMCL. In each scenario, we also see that these patterns occur for either form of prior for $\gamma_0$, although, on average, the elicited $\gamma_0$ prior consistently produces a roughly 0.3\% smaller BMCL than the objective $\gamma_0$ prior.

Table \ref{tabl:t2} gives the values for $|D(q)|$ under each scenario. We also include a relative error, measuring the maximum change in \underline{$\xi$}$_{\mbox{\tiny 100BMR}}(\epsilon)$ compared to \underline{$\xi$}$_{\mbox{\tiny 100BMR}}(0)$:
\[\delta_{\mbox{\tiny 100BMR}}=\frac{\mbox{\underline{$\xi$}$_{\mbox{\tiny 100BMR}}$}(0)-\min\left\{\mbox{\underline{$\xi$}$_{\mbox{\tiny 100BMR}}$}(\epsilon)\right\}}{\mbox{\underline{$\xi$}$_{\mbox{\tiny 100BMR}}$}(0)}.\]
The relative errors for Scenario 1 are all less than 1\%, suggesting little overall decrease in \underline{$\xi$}$_{\mbox{\tiny 100BMR}}$ as $\epsilon$ increased from 0 to 1. For Scenarios 2 and 3, the relative errors are all approximately 4\% (also noted above). Although non-trivial, such small changes suggests only minor sensitivity of the base priors to the contaminating priors, at least for these data. Values of $|D(q)|$ for Scenario 1 and 3 are all very close to 0, suggesting tiny instantaneous change of \underline{$\xi$}$_{\mbox{\tiny 100BMR}}$ at $\epsilon=0$. These are consistent with the graphical patterns seen in Figure \ref{fig:f4}. By contrast, for Scenario 2 $|D(q)|$ is roughly 1 to 3 orders of magnitudes higher than the other scenarios. This is again consistent with the patterns in Figure \ref{fig:f4}.

  \vspace{ 12 pt }
\linespread{1.2}
\begin{table}[htbp]
\begin{footnotesize}
\caption{Relative errors, $\delta_{\mbox{\tiny 100BMR}}$, and instantaneous change measure, $|D(q)|$, for each prior contamination scenario (see text) with the cumene carcinogenicity data in Table \ref{tabl:data}.}\label{tabl:t2}
\renewcommand{\arraystretch}{1.25} 
\begin{center}
\begin{tabular}{cccccc}
\hline
&\multicolumn{2}{c}{Objective beta prior}&&\multicolumn{2}{c}{Elicited beta prior}\\
\cline{2-3}
\cline{5-6}
Scenario & $\delta_{\mbox{\tiny 100BMR}}$&$|D(q)|$ &&$\delta_{\mbox{\tiny 100BMR}}$&$|D(q)|$\\
  \hline
 1& $7.679\times10^{-3}$ &$1.134\times10^{-4}$ && $5.287\times10^{-3}$ &$5.632\times10^{-6}$\\
 2& $3.767\times10^{-2}$ &$1.905\times10^{-3}$ && $3.645\times10^{-2}$ &$1.839\times10^{-3}$\\
 3& $4.477\times10^{-2}$ &$4.536\times10^{-5}$ && $4.396\times10^{-2}$&$4.174\times10^{-5}$\\
   \hline
\end{tabular}
\end{center}
\end{footnotesize}
\end{table}

\subsection{Posterior Extra Risks}\label{sec:pstER}
\begin{figure}[htbp]
\centering
\includegraphics[scale=0.3]{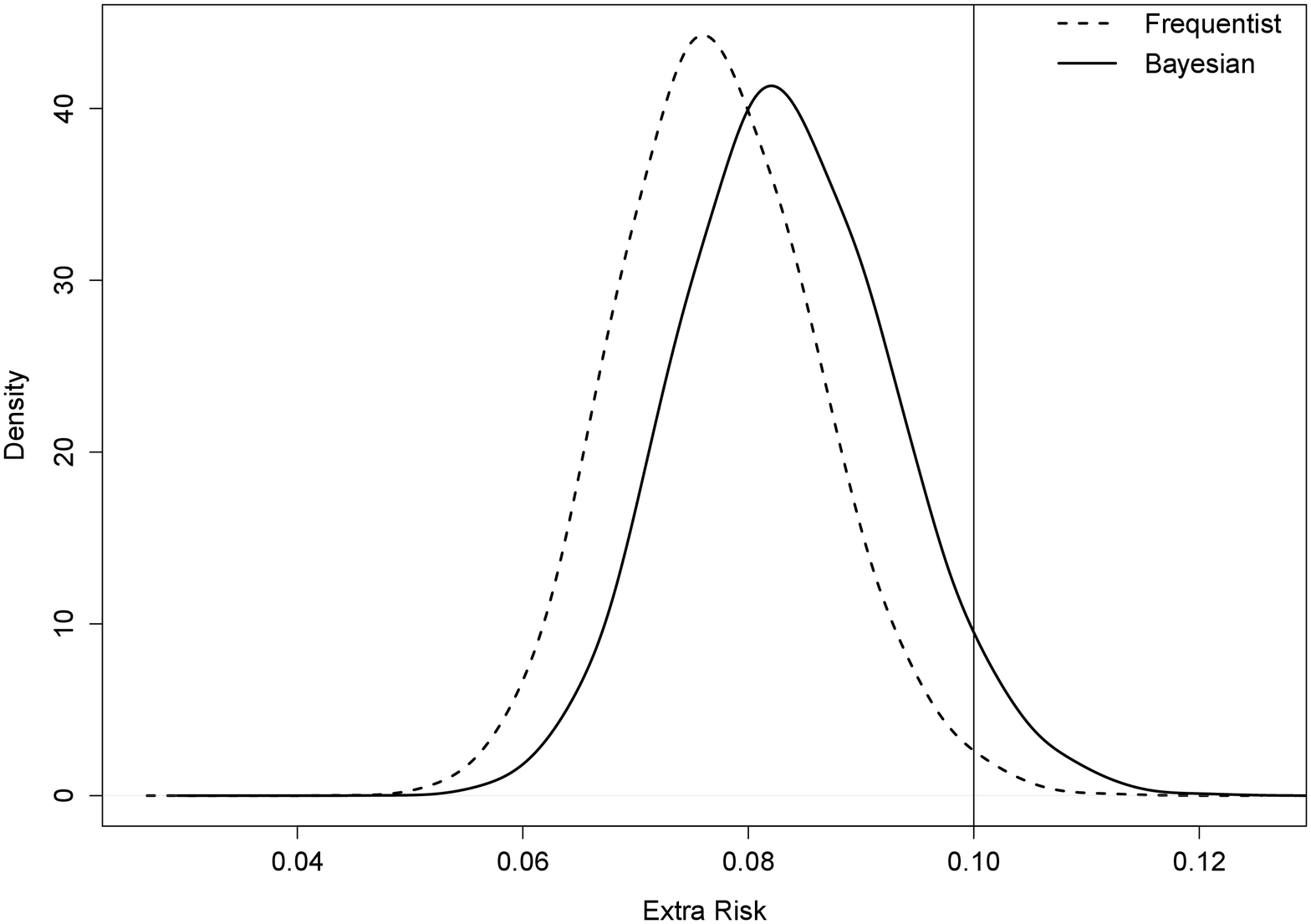}
\caption{Kernel density estimators of posterior distribution for extra risk $R_E(d)$, where $d$ is taken as the Bayesian BMCL (\solidrule) and as the frequentist BMCL (\dashedrule) in Example 1. Vertical bar indicates the BMR at 0.10.}\label{fig:DensiER}
\end{figure}

\indent\indent The posterior extra risks are further compared, according to the description in \S\ref{sec:ExtraRiskPosterior}. For a comparison, we fix two crucial dose values: our 95\% Bayesian BMCL of $d =$ \underline{$\xi$}$_{10} = 14.752$ ppm and the frequentist 95\% BMCL of $d =$ \underline{$\xi$}$_{10}= 13.618$ ppm. Figure \ref{fig:DensiER} presents kernel density estimates of the posterior extra risk at each $d$. Both density estimates appear roughly symmetric and lie fairly close to each other. As expected, since it is taken at a higher level of dose, the posterior density at the Bayesian BMCL locates farther up the extra risk scale: the mean extra risk at the Bayesian BMCL is 0.083 while the mean extra risk at the frequentist BMCL is 0.077. This reiterates the greater conservatism of the frequentist estimates with these data.

The standard deviation of the posterior extra risk at the Bayesian BMCL is 0.0096 while the standard deviation of the posterior extra risk at the frequentist BMCL is 0.0090. Again, these values are comparable; the frequentist BMCL is slightly less variable. The 95th percentile of the extra risks at the Bayesian BMCL is exactly 0.1, as expected; the 95th percentile of the extra risks at the frequentist BMCL is 0.093 which again illustrates the conservatism of the frequentist BMCL.

\begin{figure}[htbp]
\centering
\includegraphics[scale=0.3]{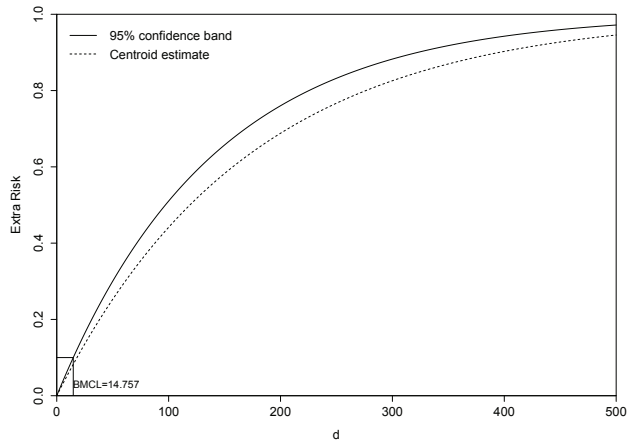}
\caption{95\% posterior credible band (\solidrule) and centroid estimate (\dashedrule) for extra risk function $R_E(d)$ in Example 1. Bayesian BMCL at BMR = 0.1 also marked, via inversion of the 95\% band.} \label{fig:ERband}
\end{figure}

Figure \ref{fig:ERband} presents a 95\% posterior credible band for the extra risk, using the method described in \S\ref{sec:ExtraRiskPosterior}.  The centroid estimate for the extra risk is also included (dashed curve).  By construction, the corresponding dose level at BMR = 0.1 on the 95\% band coincides with the Bayesian BMCL obtained above.


\section{Discussion}\label{sec:disc}
\indent\indent Herein, we consider a Bayesian approach for estimating benchmark doses (BMDs) in quantitative risk analysis.  Placing emphasis on environmental carcinogenicity assessment, our method estimates the BMD via meaningfully reparameterized quantal-response models. Prior information for the parameters is incorporated through an elicitation process, although use of objective priors is also considered. Due to the complexity of the reparameterized models, the joint posterior distribution for the unknown parameters is approximated via Monte Carlo methods, using a computationally intensive but stable adaptive Metropolis (AM) algorithm. A lower Bayesian credible limit (BMDL) is estimated as the lower percentile of the AM sample. Environmental risk analysts can employ this Bayesian approach to construct inferences on the BMD/BMDL by incorporating expert prior knowledge for the model parameters.

Of course, some caveats and qualifications are in order. Objective prior specification can be approached via many strategies when elicitation breaks down.  We have assumed independent $IG(0.001, 0.001)$ and $Beta\big(\frac12,\frac12\big)$ priors for $\xi$ and $\gamma_0$, respectively. Other forms may be pertinent, however, and we acknowledge the possible subjectivity that these choices bring to our model.


Another important issue concerns potential \emph{model uncertainty}. Our representation for the risk function was based on the popular quantal linear model. Many other forms, some quite complex, can be chosen to model $R(d)$, however.
The logistic form $R(d) =  1/(1+e^{-\beta_0 - \beta_1d})$ from \S\ref{sec:Parametric BMD} is a highly popular alternative \citep{foro07,shsm12}. Or, the quantal linear model can be extended into a `two-stage' model, $R(d) = 1 - \exp(-\beta_0 - \beta_1 d - \beta_2 d^2)$, where $\beta_j \ge 0 \; \forall j=0,1,2$.
Indeed, it is straightforward to employ our reparameterization strategy with both these models: all aspects of the Bayesian estimation scheme in \S\ref{sec:BBA} remain valid, including our elicitation strategies for the priors and the AM approach for building the joint posterior. [The three $\beta_j$s in the two-stage model require incorporation of an additional, reparameterized, nuisance parameter.  Details are given in \citet{fang14}.]

For instance, if we apply the logistic model to the cumene carcinogenicity data in Table \ref{tabl:data}, the BMC estimates on the original ppm scale are $\hat{\xi}_{10}$=42.946 ppm (if using the sample posterior median) or $\hat{\xi}_{10}$=40.892 ppm (if using the sample posterior lower tercile). The 95\% BMCL is \underline{$\xi$}$_{10}$=35.599 ppm.
These values contrast greatly with those from the quantal linear model in \S\ref{sec:exM3 result}, leading one to ask which is more appropriate---a classic example of model uncertainty.  To compare the two fits, we turn to the Bayes Factor ($BF$) \citep{kara95}: the $BF$ comparing the quantal linear model to the logistic model is calculated as the ratio of corresponding marginal likelihoods:
\begin{equation} \label{eq:BF}
BF = \frac{m_{\text{ql}}(\boldsymbol Y)}{m_{\text{lo}}(\boldsymbol Y)} \, ,
\end{equation}
where $m_{\text{ql}}(\boldsymbol Y)$ denotes the marginal likelihood using the quantal linear model and $m_{\text{lo}}(\boldsymbol Y)$ denotes the marginal likelihood using the logistic model. We approximate both marginal likelihoods using the geometric estimator and bridge sampling method from \S\ref{sec:epsilon contam}, producing $BF = 518.3$ for the cumene data. Following \citet[\S3.2]{kara95}, since this $BF$ exceeds 150 we say there is `very strong' evidence that the quantal linear model fits the data better than the logistic model.  (This is, perhaps, not surprising: the concave response observed in Figure \ref{fig:f3} would be difficult for a logistic model to match, compared to the always-concave quantal linear form.)

Clearly, model adequacy is an important issue in benchmark risk analysis \citep{wepi12}.  It has received limited attention, however, particularly from a Bayesian perspective \citep{shsm12}.  We are expanding our Bayesian BMD estimation approach to consider other popular quantal-response models, and to deal with model adequacy concerns. We hope to report on this in a future manuscript.

\section*{Acknowledgements}
Thanks are due Drs.~Anton Westveld, D.~Dean Billheimer, Susan J.~Simmons, and Cuixian (Tracy) Chen
for their helpful suggestions during the preparation of this material.  The results represent a portion of the first author's Ph.D.~dissertation with the University of Arizona Graduate Interdisciplinary Program in Statistics.

\section*{Appendix}
\indent \indent In this appendix we summarize the empirical screen employed to mark quantal data sets as dose-response `data failures.'  Over increasing doses $0=d_1<\ldots<d_i<\ldots<d_m$ $(i=1, \ldots m)$, we first calculate the empirical extra risks:
\[\tilde{R}_E(d_i)=\frac{\frac{Y_i}{N_i}-\frac{Y_1}{N_1}}{1-\frac{Y_1}{N_1}}.\]
We then connect each point $\big(d_i, \tilde{R}_E(d_i)\big)$ to the origin and find the largest slope, $S_{\scriptsize \mbox{max}}$, among all these $m-1$ rays. If $S_{\scriptsize \mbox{max}}\le0$, no increasing trend is evidenced and we mark this as a `data failure'. Notice that we do not perform a formal trend test to detect a decreasing trend \citep{whba09a}, but we do require that at least one estimated risk for some $d_i$ ($i>1$) is higher than the estimated background risk. 

\nocite*
\renewcommand\bibfont{\small}



\begin{thebibliography}{}
\begin{small}

\bibitem[\protect\citename{Andrieu and Thoms, }2008]{anth08} Andrieu, C. and Thoms, J. (2008). \newblock{ A tutorial on adaptive MCMC.} \newblock{{\em Statistics and Computing}} {\bf 18}, 343--383.


\bibitem[\protect\citename{Babb {\em et~al.}, }1998]{baro98} Babb, J., Rogatko, A., and Zacks, S. (1998). \newblock{ Cancer phase I clinical trials: efficient dose escalation with overdose control.} \textit{Statistics in Medicine} \textbf{17}, 1103--1120.






\bibitem[\protect\citename{Buckley {\em et~al.}, }2009]{buck09} Buckley, B.~E., Piegorsch, W.~W. and West, R. W. (2009). \newblock{ Confidence limits on one-stage model parameters in benchmark risk assessment.} \newblock {\em Environmental and Ecological Statistics} {\bf 16}, 53--62.



\bibitem[\protect\citename{Casella and Berger, }2002]{cabe02} Casella, G. and Berger, R.~L. (2002).  \newblock {\em Statistical Inference}, 2nd edn. \newblock Pacific Grove, CA: Duxbury.


\bibitem[\protect\citename{Christensen {\em et~al.}, }2011]{chjo11} Christensen, R., Johnson, W.~O., Branscum, A.~J. and Hanson, T.~E. (2011). \newblock {\em Bayesian Ideas and Data Analysis: An Introduction for Scientists and Statisticians}. \newblock Boca Raton, FL: Chapman \& Hall/CRC Press.

\bibitem[\protect\citename{Coull {\em et~al.}, }2003]{come03} Coull, B.~A., Mezzetti, M. and Ryan, L.~M. (2003). \newblock{ A Bayesian hierarchical model for risk assessment of methylmercury.} \textit{Journal of Agricultural, Biological, and Environmental Statistics} \textbf{8}, 253--270.


\bibitem[\protect\citename{Crump, }1984]{crum84} Crump, K.~S. (1984). \newblock{ A new method for determining allowable daily intake.} \newblock {\em Fundamental and Applied Toxicology} {\bf 4}, 854--871.

\bibitem[\protect\citename{Crump, }1995]{crum95} Crump, K.~S. (1995). \newblock{ Calculation of benchmark doses from continuous data.} \newblock {\em Risk Analysis} {\bf 15}, 79--89.


\bibitem[\protect\citename{Davis {\em et~al.}, }2012]{davi12} Davis, J.~A., Gift, J.~S. and Zhao, Q.~J. (2012). \newblock{ Introduction to benchmark dose methods and U.S. EPA's Benchmark Dose Software (BMDS) version 2.1.1.} \newblock {\em Toxicology and Applied Pharmacology} {\bf 254}, 181--191.

\bibitem[\protect\citename{European Union, }2003]{eu03} European Union (2003). \textit{ Technical Guidance Document (TGD) on Risk Assessment of Chemical Substances following European Regulations and Directives, Parts I-IV}. \newblock{ Technical Report number EUR 20418 EN/1-4.} \newblock Ispra, Italy: European Chemicals Bureau (ECB).

\bibitem[\protect\citename{Fang, }2014]{fang14} Fang, Q. (2014). \textit{Hierarchical Bayesian Benchmark Risk Analysis}. \newblock {Ph.D. thesis, Interdisciplinary Program in Statistics, University of Arizona, Tucson, AZ.}


\bibitem[\protect\citename{Foronda {\em et~al.}, }2007]{foro07} Foronda, N.~M., Fowles, J., Smith, N., Taylor, M. and Temple, W. (2007).  \newblock{ A benchmark dose analysis for sodium monofluoroacetate (1080) using dichotomous toxicity data.} \newblock {\em Regulatory Toxicology and Pharmacology} {\bf 47}, 84--89.





\bibitem[\protect\citename{Gelman, }2012]{gelm12} Gelman, A. (2012). \newblock{ Posterior distribution.} \newblock{In El-Sharaawi, A.~H. and Piegorsch, W.~W. (eds.)}, \textit{Encyclopedia of Environmetrics}, 2nd edn., \textbf{4}, 2020--2021. Chichester: John Wiley \& Sons.







\bibitem[\protect\citename{Guha {\em et~al.}, }2013]{guro13} Guha,~N., Roy,~A., Kopylev,~L., Fox,~J., Spassova,~N., and White,~P. (2013). \newblock{ Nonparametric Bayesian methods for benchmark dose estimation.} \textit{Risk Analysis} {\bf 9}, 1608--1619.  

\bibitem[\protect\citename{Held, }2004]{held04} Held, L. (2004). \newblock{ Simultaneous inference in risk assessment; a Bayesian perspective.} \newblock {In Antoch, J.~(ed.), {\em COMPSTAT 2004.  Proceedings in Computational Statistics}, 213--222.} \newblock{Heidelberg: Physica-Verlag.}






\bibitem[\protect\citename{Kass and Raftery, }1995]{kara95} Kass, R.~E. and Raftery, A.~E. (1995). \newblock{ Bayes Factor.} \textit {Journal of the American Statistical Association} \textbf{90}, 773--795.



\bibitem[\protect\citename{Kuhnert, }2011]{kuhn11} Kuhnert, P.~M. (2011). \newblock{ Four case studies in using expert opinion to inform priors.} {\em Environmetrics} \textbf{22}, 662--674.

\bibitem[\protect\citename{Kvam and Vidakovic, }2007]{kvvi07} Kvam, P.~H. and Vidakovic, B. (2007). {\em Nonparametric Statistics with Applications to Science and Engineering}.  Hoboken, NJ: John Wiley \& Sons.

\bibitem[\protect\citename{Lambert {\em et~al.}, }2005]{lamb05} Lambert, P.~C., Sutton, A.~J., Burton, P.~R., Abrams, K.~R. and Jones, D.~R. (2005). \newblock{ How vague is vague? A simulation study of the impact of the use of vague prior distributions in MCMC using WinBUGS.} \newblock {\em Statistics in Medicine} {\bf 24}, 2401--2428.


\bibitem[\protect\citename{Lopes and West, }2004]{lope04} Lopes, H.~F. and West, M. (2004). \newblock{ Bayesian model assessment in factor analysis.} \newblock{\em Statistica Sinica} {\bf 14}, 41--67.

\bibitem[\protect\citename{Meng and Wong, }1996]{meng96} Meng, X. and Wong W.~H. (1996). \newblock{ Simulating ratios of normalizing constants via
a simple identity: A theoretical exploration.} \newblock{\em Statistica Sinica} {\bf 6}, 831--860.


\bibitem[\protect\citename{Morales {\em et~al.}, }2006]{moib06} Morales, K.~H., Ibrahim, J.~G., Chen, C.-J. and Ryan, L.~M. (2006).  \newblock{ Bayesian model averaging with applications to benchmark dose estimation for arsenic in drinking water.} \newblock {\em Journal of the American Statistical Association} {\bf 101}, 9--17.



\bibitem[\protect\citename{Naufal {\em et~al.}, }2009]{nauf09} Naufal, Z., Kathman, S. and Wilson, C. (2009). \newblock{ Bayesian derivation of an oral cancer slope factor distribution for 4-(methylnitrosamino)-1-(3-pyridyl)-1-butanone (NNK).} \newblock {\em Regulatory Toxicology and Pharmacology} {\bf 55}, 69--75.

\bibitem[\protect\citename{Nitcheva {\em et~al.}, }2007]{nitc07} Nitcheva, D.~K., Piegorsch, W.~W. and West, R.~W. (2007). \newblock{ On use of the multistage dose-response model for assessing laboratory animal carcinogenicity.} \newblock {\em Regulatory Toxicology and Pharmacology} {\bf 48}, 135--147.


\bibitem[\protect\citename{OECD, }2006]{oecd06} OECD (2006). \textit{ Current Approaches in the Statistical Analysis of Ecotoxicity Data: A Guidance to Application, Series on Testing and Assessment No. 54}. \newblock Paris: Environment Directorate, Organisation For Economic Co-Operation and Development.

\bibitem[\protect\citename{OECD, }2008]{oecd08} OECD (2008). \textit{ Draft Guidance Document on the Performance of Chronic Toxicity and Carcinogenicity Studies, Supporting TG 451, 452 and 453}. \newblock Paris: Organisation For Economic Co-Operation and Development.

\bibitem[\protect\citename{O'Hagan, }1994]{ohag94} O'Hagan, A. (1994). \textit{Kendall's Advanced Theory of Statistics, Volume 2B, Bayesian Inference}, 2nd edn. London: Edward Arnold.

\bibitem[\protect\citename{O'Hagan {\em et~al.}, }2006]{ohbu06} O'Hagan, A., Buck, C.~E., Daneshkhah, A., Eiser, J.~R., Garthwaite, P.~H., Jenkinson, D.~J., Oakley, J.~E. and Rakow, T. (2006). \textit{Uncertain Judgements: Eliciting Experts' Probabilities}. Chichester: John Wiley \& Sons.


\bibitem[\protect\citename{Parham and Portier, }2005]{papo05} Parham, F. and Portier C.~J. (2005). \newblock{ Benchmark dose approach.} \newblock {In Edler, L. and Kitsos, C. (eds.), {\em Recent Advances in Quantitative Methods in Cancer and Human Health Risk Assessment}, 239--254.} \newblock{Chichester: John Wiley \& Sons.}


\bibitem[\protect\citename{Piegorsch and Bailer, }2005]{piba05} Piegorsch, W.~W. and Bailer, A.~J. (2005). \newblock {\em Analyzing Environmental Data}. \newblock Chichester: John Wiley \& Sons.




\bibitem[\protect\citename{{R Development Core Team}, }2012]{r12} {R Development Core Team}. (2012). \newblock {\em R: A Language and Environment for Statistical Computing}. \newblock Vienna, Austria: R Foundation for Statistical Computing. \newblock {ISBN} 3-900051-07-0.




\bibitem[\protect\citename{Shao, }2012]{shao12} Shao, K. (2012). \newblock{  A comparison of three methods for integrating historical information for Bayesian model averaged benchmark dose estimation.}  \textit{Environmental Toxicology and Pharmacology} \textbf{34}, 288--296.

\bibitem[\protect\citename{Shao and Small, }2011]{shsm11} Shao, K. and Small, M.~J. (2011). \newblock{ Potential uncertainty reduction in model-averaged benchmark dose estimates informed by an additional dose study.} \newblock {\em Risk Analysis} {\bf 31}, 1561--1575.

\bibitem[\protect\citename{Shao and Small, }2012]{shsm12} Shao, K. and Small, M.~J. (2012). \newblock{ Statistical evaluation of toxicological experimental design for Bayesian model averaged benchmark dose estimation with dichotomous data.} \newblock {\em Human and Ecological Risk Assessment} {\bf 18}, 1096--1119.


\bibitem[\protect\citename{Stern, }2008]{ster08} Stern, A.~H. (2008). \newblock{ Environmental health risk assessment.} \newblock {In Melnick, E.~L. and Everitt, B.~S. (eds.), {\em Encyclopedia of Quantitative Risk Analysis and Assessment} {\bf 2}, 580--589. \newblock Chichester: John Wiley \& Sons.}


\bibitem[\protect\citename{U.S.~EPA, }2012]{epa12} U.S.~EPA (2012). \newblock{ \em Benchmark Dose Technical Guidance Document}. \newblock{ Technical Report number EPA/100/R-12/001.} \newblock Washington, DC: U.S. Environmental Protection Agency.


\bibitem[\protect\citename{U.S.~General Accounting Office, }2001]{gao01} U.S.~General Accounting Office (2001). \textit{ Chemical Risk Assessment.  Selected Federal Agencies' Procedures, Assumptions, and Policies}. \newblock{ Report to Congressional Requesters number GAO-01-810.} \newblock Washington, DC: U.S.~General Accounting Office.

\bibitem[\protect\citename{U.S.~NTP, }2009]{tr542} U.S.~National Toxicology Program (2009). \newblock{ \em Toxicology and Carcinogenesis Studies of Cumene (CAS NO. 98-82-8) in F344/N Rats and B6C3F$_1$ Mice}. \newblock{ Technical Report number 542.} \newblock Research Triangle Park, NC: U.S. Department of Health and Human Services, Public Health Service.






\bibitem[\protect\citename{West {\em et~al.}, }2012]{wepi12} West, R.~W., Piegorsch, W.~W., Pe\~na, E.~A., An, L., Wu, W., Wickens, A.~A., Xiong, H., and Chen, W. (2012). \newblock{ The impact of model uncertainty on benchmark dose estimation.} \newblock {\em Environmetrics} {\bf 23}, 706-716.

\bibitem[\protect\citename{Wheeler and Bailer, }2009a]{whba09a} Wheeler, M.~W. and Bailer, A.~J. (2009a). \newblock{ Comparing model averaging with other model selection strategies for benchmark dose estimation.} \newblock{\em Environmental and Ecological Statistics} {\bf 16}, 37--51.

\bibitem[\protect\citename{Wheeler and Bailer, }2009b]{whba09b} Wheeler, M.~W. and Bailer, A.~J. (2009b). \newblock{ Benchmark dose estimation incorporating multiple data sources.} \newblock{\em Risk Analysis} {\bf 29}, 249--256.

\bibitem[\protect\citename{Wheeler and Bailer, }2012]{whba12} Wheeler, M.~W. and Bailer, A.~J. (2012). \newblock{ Monotonic Bayesian semiparametric benchmark dose analysis.} \newblock{\em Risk Analysis} {\bf 32}, 1207--1218.


\end{small}
\end{thebibliography}
\end{document}


\begin{center}
\noindent \textbf{\large{Supplement to `Bayesian Benchmark Dose Analysis'}}

\vspace{8 pt}
\renewcommand\thefootnote{\fnsymbol{footnote}}
\noindent \textbf{Qijun Fang}$^{\textrm{a}}$\symbolfootnote[2]{Email address: qijunf@email.arizona.edu; Corresponding author}, \textbf{Walter W.~Piegorsch}$^{\textrm{a,b}}$, and \textbf{Katherine Y.~Barnes}$^{\textrm{a,c}}$
\\
\noindent $^{\textrm{a}}$\textit{Program in Statistics}, $^{\textrm{b}}$\textit{BIO5 Institute}, and $^{\textrm{c}}$\textit{James E.~Rogers College of Law}\\
\textit{University of Arizona, Tucson, AZ, USA}\\

\end{center}
\vspace{1 pc}

\noindent This supplementary document provides supporting material for \emph{Bayesian Benchmark Dose Analysis} by Qijun Fang and Walter W. Piegorsch (the `main document').  The various sections below address a variety of supplemental/supporting topics, and are not intended to flow naturally between each other.  They are, however, presented in roughly the same order in which their counterpart topics appear in the main document.  As there, we denote the Benchmark Dose (BMD) target parameter as $\xi$ and the background response probability nuisance parameter as $R(0) = \gamma_0$.

\section{Finding prior parameters from elicited quantiles}\label{sec:suppl1}

\indent \indent We give here technical aspects on derivation of the prior parameters $\alpha$, $\beta$, $\psi$, and $\omega$ for the prior densities $\xi \sim \mbox{\textit{IG}}(\alpha, \beta)$ and $\gamma_0 \sim \mbox{\textit{Beta}}(\psi, \omega)$ used in the main document's hierarchical model.  The elicited lower quartiles and medians $Q_{1\xi}, Q_{2\xi}$ for $\xi$, and $Q_{1\gamma}, Q_{2\gamma}$ for $\gamma_0$, respectively, are assumed given from domain expert(s) input.

Start with the elicitation for $\xi$:  by definition, $Q_{1\xi}$ and $Q_{2\xi}$ satisfy
\begin{equation*}\label{eq:xi.1}
\int_0^{Q_{j\xi}}\frac{\beta^\alpha}{\Gamma(\alpha)} \xi^{(-\alpha-1)} \exp\left\{-\frac{\beta}{\xi}\right\}d\xi = \frac{j}{4},
\end{equation*}
($j = 1,2$), where the integrand is the IG p.d.f.  This establishes a system of non-linear equations for $\alpha$ and $\beta$.  Unfortunately, no closed-form solution exists for the system, and so we turn to numerical methods.  We employ a gradient method discussed by \cite{barz88} and implemented in the \verb|R| statistical environment \citep{r12} via the \verb|BB| package \citep{vara09}. This method employs iterative updating until half the $L_2$ norm of the system at the proposed solution is smaller than $10^{-10}$.

Similarly, for $\gamma_0$ the elicited quartiles $Q_{1\gamma}$ and $Q_{2\gamma}$ satisfy
\begin{equation*}\label{eq:xi.2}
\int_0^{Q_{j\gamma}}\frac{\Gamma(\psi+\omega)}{\Gamma(\psi)\Gamma(\omega)} \gamma_0^{\psi-1}(1-\gamma_0)^{\omega-1}d\gamma_0 = \frac{j}{4},
\end{equation*}
($j = 1,2$), now applying the p.d.f.~for the beta prior.  Here again, the resulting system of non-linear equations possesses no closed-form solution, so we appeal to the \verb|BB| package in \verb|R|.

The resulting iterative solutions to these two systems of equations produce values for the prior parameters $\alpha$, $\beta$, $\psi$, and $\omega$ which are then employed in our hierarchical model for the BMD.

\section{The adaptive Metropolis algorithm}\label{sec:suppl2}

\indent \indent The posterior p.d.f.~under our hierarchical model is
\begin{eqnarray}\label{eq:post2}
\nonumber \pi(\xi, \gamma_0|{\boldsymbol Y})&=&\frac{f({\boldsymbol Y}|\xi, \gamma_0)\pi(\xi|\alpha,\beta)\pi(\gamma_0|\psi,\omega)}{m(\boldsymbol Y)}\\
 \\
\nonumber &=&\frac{\prod_{i=1}^n{R(d_i)}^{Y_i}{[1-R(d_i)]}^{N_i-Y_i}}{m(\boldsymbol Y)}
\frac{\beta^\alpha e^{-\beta/\xi}}{\Gamma(\alpha)\xi^{\alpha+1}}
\frac{\Gamma(\psi+\omega)}{\Gamma(\psi)\Gamma(\omega)}\gamma_0^{\psi-1}(1-\gamma_0)^{\omega-1},
\end{eqnarray}
where
\[m(\boldsymbol Y) = \int_0^\infty\int_0^1 f({\boldsymbol Y}|\xi, \gamma_0)\pi(\xi|\alpha,\beta)\pi(\gamma_0|\psi,\omega)d\gamma_0d\xi.\]
By employing proper priors, this is guaranteed to be integrable \citep[\S 5.3]{geca04}, which helps motivate our preference for the IG and beta prior assumptions on $\xi$ and $\gamma_0$, respectively.  Nonetheless, the posterior is intractable, so we turn to approximation by computer.  Markov chain Monte Carlo (McMC) approaches have been applied with great success to models such as this \citep{roca11}, although the complexity of \eqref{eq:post2} prevents us from applying well-known McMC algorithms such as Gibbs sampling \citep{gelf00}. Instead, we move to Metropolis-Hastings approaches \citep[Ch.~7]{roca04} and in particular consider \emph{adaptive} Metropolis (AM) algorithms.  These tune the variance of the underlying Metropolis proposal density adaptively when generating ongoing draws of the chain. From our experience with a variety of such methods, we favor a global AM procedure with componentwise adaptive scaling described by \citet{anth08}. Their `AM6' algorithm uses all previous iterations in the chain to update the variance-covariance matrix of the current bivariate proposal density.  The operation is global in the sense that both the $\xi$ and $\gamma_0$ parameters are updated simultaneously at each $k$th bivariate draw; $k=1,\ldots,K$. It is also componentwise in the sense that the updating step sizes for $\xi$ and $\gamma_0$ are controlled by separate adaptive scaling parameters.  See \cite{anth08} for further details.

To begin the AM chain, `starting' points for $\xi$ and $\gamma_0$ must be chosen.  Rather than applying a random initialization, we take the starting point for $\gamma_0$ as $\frac{Y_1+0.25}{N_1+0.5}$, i.e., the estimated nonresponse probability with a slight shrinkage term included to pull away from the zero boundary.

The starting point for $\xi$ was obtained by first treating $R_E(d)=S_{\scriptsize \mbox{max}}d$ as an empirical approximation to the extra risk function and then solving for $d$ in the equation $\mbox{BMR}=S_{\scriptsize \mbox{max}}d$.  Here, $S_{\scriptsize \mbox{max}}$ is the maximum empirical origin-to-response slope described in the main document's Appendix.  We then find the starting point for $\xi$ as simply BMR/$S_{\scriptsize \mbox{max}}$.

The AM6 algorithm employs a bivariate normal proposal distribution. \cite{anth08} recommend $\frac{2.38^2}{2}\mbox{\bf I}$ as an initial variance-covariance matrix, where ${\bf I}$ is the $2\times 2$ identity matrix. At each subsequent iteration, the variance-covariance matrix of the normal proposal density is then updated based on the previous iteration's variance-covariance matrix and on the current iteration's adaptive scaling parameters for $\xi$ and $\gamma_0$. See \cite{anth08} for details.

\section{Monte Carlo `burn-in' diagnostics}\label{sec:suppl3}

\indent \indent
To account for potential instability in the early portions of the bivariate chain, we include a `burn-in' for the Monte Carlo sample \citep[\S11.6]{geca04}.  We couple this with the larger question of how to assess the chain's convergence.
Many approaches exist for diagnosing McMC convergence \citep{coca96}, and indeed, the issue is a topic of ongoing debate \citep[\S12.2]{roca04}.
We mimic a method due to \citet{gewe92}, where early portions of the chain are sub-sampled and compared against latter portions to determine where the larger sample of $K$ draws begins to exhibit stable performance.  To approximate independence between the two sub-samples, we bifurcate the chain by a gap of no less than $\frac{K}{5}$ consecutive draws.

For summary diagnostics, we calculate the difference in arithmetic means between the two bifurcated sub-samples, and divide each difference by its standard error to produce a $Z$-statistic.  The approximate standard error is taken as the square root of the sum of the variances of each mean.  This is done separately for both the $\xi_k$ and $\gamma_{0k}$ components, $k=1,\ldots,K$.  (For sake of simplicity, we make no correction for multiplicity.)  To adjust for possible autocorrelation within each sub-sample, the individual variances are based on estimated spectral densities at frequency zero; see \cite{fang13} for full details.

To monitor if the pattern of association between $\xi$ and $\gamma_0$ within the larger chain also exhibits reasonable stability, we include comparison of the covariances between the two sub-samples.  As a diagnostic measure here we use the sample covariance from each sub-sample,
$$\widehat{\mbox{Cov}}(\xi, \gamma_0)=\frac{1}{L}\sum_{k=1}^L(\xi_k-\bar{\xi})(\gamma_{0k}-\bar{\gamma}_0), $$
where $\bar{\xi}$ and $\bar{\gamma}_0$ are the pertinent arithmetic means within a bifurcated sub-sample of length $L$.
To find the approximate variance of each estimated covariance we calculate the quantities
$\psi_k = (\xi_k-\bar{\xi})({\gamma_{0k}}-\bar{\gamma_0})$
across all values of $k$ in the given sub-sample and then estimate the variance of $\psi_k$ based on their estimated spectral densities at frequency zero.

Each such comparison is conducted over three individual, separated bifurcations of the full $K$-length sample: (i) the first 10\% of the chain ($L = K/10$) vs.~the final 50\% ($L = K/2$); (ii) the first 20\% of the chain ($L = K/5$) vs.~the final 50\%; and (iii) the first 30\% of the chain ($L = 0.3K$) vs.~the final 50\%. We consider an individual diagnostic comparison as a `pass' if the corresponding $Z$-statistic comparing the two bifurcated samples is less than 1.96 in absolute value.  To `pass' the full diagnostic at each bifurcation, all three measures---mean of $\xi$, mean of $\gamma_0$, and covariance of $\{\xi$,$\gamma_0\}$---must individually pass.

The comparisons are performed in sequential order:  if the 10\%-vs.-50\% diagnostic fails, then the 20\%-vs.-50\% diagnostic is conducted. If this fails, we move to the 30\%-vs.-50\% diagnostic. When a diagnostic passes, we view the indicated early portion of the chain as a reasonable burn-in.  For notation, we let $K_0 < K$ be the resulting index that begins the retained portion of the chain.  So, e.g., if the 10\%-vs.-50\% diagnostic fails but then the 20\%-vs.-50\% diagnostic passes, we take the first 20\% of the chain as burn-in and use the remaining 80\% of the chain as our Monte Carlo sample from $\pi(\xi, \gamma_0|{\boldsymbol Y})$.  If $K = 100,000$, as used throughout, this gives $K_0 = 20,001$.

If none of the three sequential diagnostics passes, we re-set the random-number generator's seed and reassess a new set of $K$ Monte Carlo draws.  If after five such re-starts the diagnostic continues to fail, we report an \emph{algorithm failure}.  (This did not occur very often; see the simulation results in \S\ref{sec:suppl6}, below.)

\section{Estimating marginals via bridge sampling}\label{sec:suppl4}

\indent \indent In the main document, we often have reason to calculate (an approximation to) a marginal likelihood of the form
\[m(\boldsymbol Y) = \int_0^\infty\int_0^1 f({\boldsymbol Y}|\xi, \gamma_0)\pi(\xi|\alpha,\beta)\pi(\gamma_0|\psi,\omega)d\gamma_0d\xi\]
for various choices of $\pi(\xi|\alpha,\beta)$ and/or $\pi(\gamma_0|\psi,\omega)$.
To do so, we construct a \emph{geometric estimator} from our AM sample, appealing to the `bridge sampling' method recommended by \cite{meng96} and \cite{lope04}.  To approximate the marginal likelihood, an approximation $g(\cdot)$ to the joint posterior density of $(\xi, \gamma_0)$ is first chosen. For simplicity, we took $g(\xi, \gamma_0)$ as the bivariate normal density function with mean set to the empirical mean vector and variance set to the empirical variance-covariance matrix of $\xi$ and $\gamma_0$ estimated from the AM sample. Next, a set of $K^* = K - K_0 + 1$ bivariate vectors $\big\{[\xi^*_j~ \gamma^*_{0j}]^{\mbox{\scriptsize \textit{T}}}\big\}_{j=1}^{K^*}$ are drawn from $g(\xi, \gamma_0)$. The marginal likelihood is then approximated by substituting the (retained) AM sample, $\big\{[\xi_k, \gamma_{0k}]^{\mbox{\scriptsize \textit{T}}}\big\}_{k=K_0}^K$, and the generated sample from $g(\xi, \gamma_0)$, $\big\{[\xi^*_j~ \gamma^*_{0j}]^{\mbox{\scriptsize \textit{T}}}\big\}_{j=1}^{K^*}$, into the ratio
\[\widehat{m}(\boldsymbol Y)=\frac{\frac{1}{K^*}\sum_{j=1}^{K^*}\left\{\pi\left(\xi^{*}_j|\alpha,\beta\right)\pi\left(\gamma_{0j}^{*}|\psi,\omega\right)f\left({\boldsymbol Y}|\xi^{*}_j,\gamma_{0j}^{*}\right)/g\left(\xi^{*}_j,\gamma_{0j}^{*}\right)\right\}^{1/2}}
{\frac{1}{K^*}\sum_{k=K_0}^K\left[g\left(\xi_k,\gamma_{0k}\right)/\left\{\pi\left(\xi_k|\alpha,\beta\right)\pi\left(\gamma_{0k}|\psi,\omega\right)f\left({\boldsymbol Y}|\xi_k,\gamma_{0k}\right)\right\}\right]^{1/2}},\]
where $f(\cdot)$ denotes the binomial likelihood function and $\pi(\cdot)$ denotes the pertinent prior density [e.g., $q(\xi)$ when approximating $m_q(\boldsymbol Y)$ in our $\epsilon$-contamination study].

\section{Prior elicitation with the cumene data}\label{sec:suppl5}

\indent \indent  To elicit the prior parameters for the cumene data in Example 1, we applied input from domain experts and based the prior elicitation on existing background from the toxicological literature.
We began with the BMD target parameter, $\xi$:  an oral No Observed Adverse Effect Level (NOAEL) for long-term cumene exposure in rodents was given by the EPA IRIS website (\verb|http://www.epa.gov/iris/subst/0306.htm|) as 110 mg/kg-day. Guidance for converting  the oral mg/kg-day metric to inhalation ppm (the scale used in the original data table from the main document) was available via conversion data given by the \cite{OEHHA04}:  for rodents, 1 mg/kg-day works out to roughly 2.25 ppm.  Arguing that a NOAEL is loosely equivalent to a BMD---although, see \citep{kode05} for an in-depth discussion---the median prior estimate was $Q_{2\xi} = 110$ mg/kg-day = 247.5 ppm $\approx 250$ ppm.  For the first quartile, the EPA IRIS site gave a cumene inhalation NOAEL for rodents as 435 mg/cu.m, based on a shorter, 13-week study. The shorter exposure period was suited to provide only a limited estimate on the BMD's central tendency; we translated this to use the 435 mg/cu.m 13-week NOAEL as a prior estimate of the first quantile.  For conversion to inhalation ppm, the EPA indicated that roughly 1 mg/cu.m = 0.204 ppm, so we took $Q_{1\xi} = 435$ mg/cu.m = 88.75 ppm $\approx 90$ ppm. We then divided the ppm dose by 500 ppm to standardize the $\xi$-scale to $0\le \xi \le1$. This produced prior first quartile and median estimates for $\pi(\xi|\alpha, \beta)$ as 0.18 and 0.5, respectively.

For $\gamma_0$, historical control data on alveolar/bronchiolar tumors in B6C3F$_1$ mice were given by \citet[Table D3a]{tr542}. These provide direct prior information on the background risk.  The historical data gave a median background response of $Q_{2\gamma} = 0.08$ and a lower quartile of $Q_{1\gamma} = 0.04$.

These quartile-based elicitations were then used to construct the prior densities $\xi \sim \mbox{\textit{IG}}(\alpha, \beta)$ and $\gamma_0 \sim \mbox{\textit{Beta}}(\psi, \omega)$ as described in \S\ref{sec:suppl1}, above.

\section{Trace plot and convergence diagnostics for cumene example}\label{sec:suppl6}
\begin{figure}[htbp]
\centering
\includegraphics[scale=0.3]{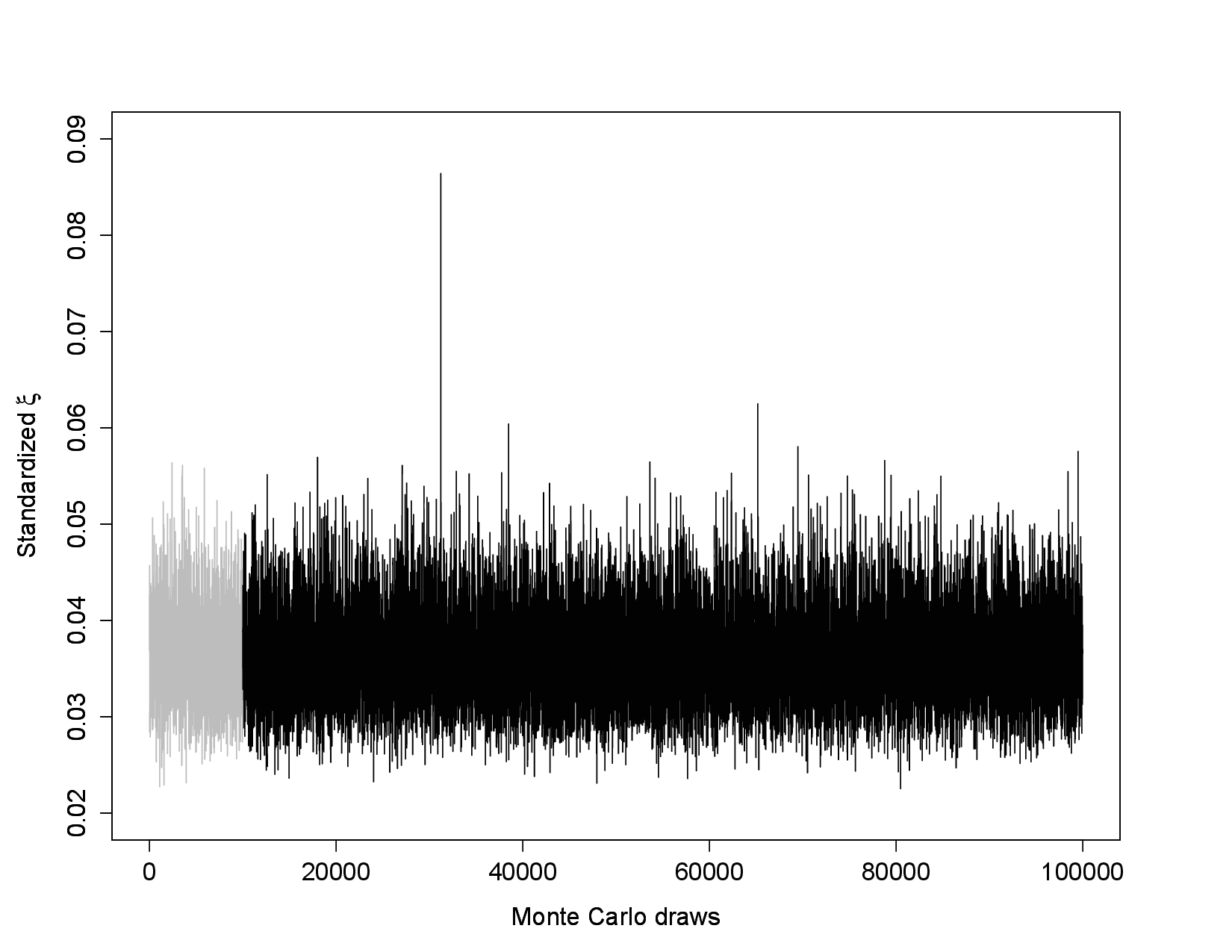}
\caption{Trace plot for $\xi$ in Example 1. Dose scale ($\xi$) is standardized to unit length after dividing by the highest dose in the data set. First 10,000 Monte Carlo draws in grey indicate burn-in period.}\label{fig:f1}
\end{figure}

\indent\indent In the cumene example, an AM approach was applied to approximate the posterior for $\xi$ and $\gamma_0$. At an initial chain size of $100,000$, Figure \ref{fig:f1} displays the full trace plot for the $\xi$ component of this AM sample. The plot exhibits stable performance and fairly fast mixing.  Our convergence diagnostic procedure recommended an initial burn-in of $10,000$ draws.  The remaining $90,000$ draws were then used as the posterior approximation for these data.

